\documentclass[aps,prl,superscriptaddress,floatfix,letterpaper,preprint]{revtex4}

\usepackage{amsmath}
\usepackage{amssymb}
\usepackage{graphicx}
\usepackage[urlcolor=blue, hyperindex, colorlinks, bookmarks=true]{hyperref}
\usepackage{natbib}
\usepackage{mathtools}%

\newcommand{\bra}[1]{\langle #1|}
\newcommand{\ket}[1]{|#1\rangle}
\newcommand{\unit}[1]{\ensuremath{\, \mathrm{#1}}}

\begin{document}
\date{August 14, 2013}
\title{Time-Reversal Symmetrization of Spontaneous Emission for High Fidelity Quantum State Transfer}

\author{Srikanth~J.~Srinivasan}
\affiliation{Department of Electrical Engineering, Princeton University, Princeton, NJ 08544, USA}
\author{Neereja~M.~Sundaresan}
\affiliation{Department of Electrical Engineering, Princeton University, Princeton, NJ 08544, USA}
\author{Darius~Sadri}
\affiliation{Department of Electrical Engineering, Princeton University, Princeton, NJ 08544, USA}
\author{Yanbing~Liu}
\affiliation{Department of Electrical Engineering, Princeton University, Princeton, NJ 08544, USA}
\author{Jay~M.~Gambetta}
\affiliation{T. J. Watson Research Center, Yorktown Heights, NY 10598 USA}
\author{Terri Yu}
\affiliation{Departments of Physics and Applied Physics, Yale University, New Haven, CT 06511, USA}
\author{S.~M.~Girvin}
\affiliation{Departments of Physics and Applied Physics, Yale University, New Haven, CT 06511, USA}
\author{Andrew~A.~Houck}
\affiliation{Department of Electrical Engineering, Princeton University, Princeton, NJ 08544, USA}

\begin{abstract}
We demonstrate the ability to control the spontaneous emission from a superconducting qubit coupled to a cavity. The time domain profile of the emitted photon is shaped into a symmetric truncated exponential. The experiment is enabled by a qubit coupled to a cavity, with a coupling strength that can be tuned in tens of nanoseconds while maintaining a constant dressed state emission frequency. Symmetrization of the photonic wave packet will enable use of photons as flying qubits for transfering the quantum state between atoms in distant cavities.
\end{abstract}

\DeclareGraphicsExtensions{.pdf}

\maketitle

\begin{figure}
\begin{center}
	\includegraphics[width=.8\textwidth]{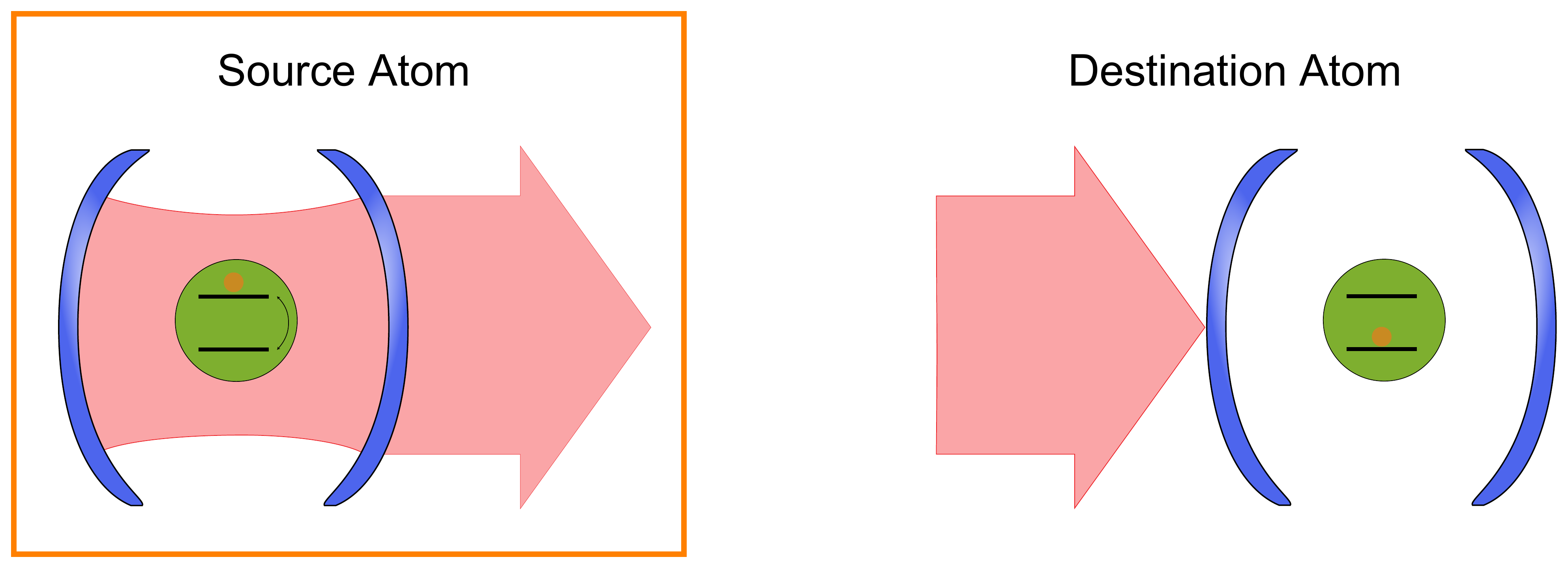}
\caption{\label{labCartoon}Schematic of the State Transfer Experiment. Transferring a quantum state starts with an atom in a cavity in some superposition of ground and excited states. The atom-cavity coupling induces the atom to encode its quantum state onto a photonic wave packet. This wave packet travels through a wave guide and impinges onto the input port of a destination cavity, where the photonic state needs to be encoded onto the atom. When the wave packet arrives at the destination, the large impedance mismatch presented by the cavity's input port reflects the pulse and state transfer fidelity suffers. One solution is to symmetrize the wave packet such that emission by the source and absorption by the destination are time-reversed processes. The orange rectangle delineates the scope of this paper; the time-domain emission profile from a superconducting qubit is symmetrized. This is achieved by modulating the coupling strength to a cavity during the photon emission.}
\end{center}
\end{figure}

Transferring information on a classical computer is empowered by the data bus. The analogue for shuttling quantum information is a more challenging problem, but can addressed by using a single quantum emitter. It enables the generation of exotic states of light such as those containing a single photon~\cite{Haroche2006}. Photons are well-suited for this task due to their long coherence lengths, and ability to encode quantum information. When an atom in some arbitrary superposition of ground and excited states relaxes, its emission can be described in the basis of zero and one photon states~\cite{Houck2007}. The exact time at which this relaxation happens is a random Poissonian process with average time $T_1$. The emission will have a wave packet shaped with a steep leading-edge front followed by an exponential tail with time constant $T_1$. If this wave packet is used in a state transfer scheme as in Fig.~\ref{labCartoon}, it will be reflected at the destination cavity's input port due to what is effectively an impedance mismatch even if the destination cavity is identical to the source cavity.

To counteract this loss in transfer fidelity, one can engineer the wave packet emitted by the source to be symmetric in time, preserving time-reversal invariance. The destination atom-cavity system properties can then be appropriately tuned to make absorption by the destination akin to the time-reversed process of emission by the source. Theoretical estimates for this technique predict near unit transfer fidelity~\cite{Cirac1997,Korotkov2011}. One approach to state transfer is to load the cavity with a single photon, and then vary the cavity leak rate to match the desired emission profile~\cite{Yin2013}. At optical frequencies, careful control of the pump laser has been used to produce single photons with a desired shape~\cite{Keller2004}. In another experiment, an electro-optic modulator has been used to modulate a single photon from a biphoton pair generated by spontaneous parametric down-conversion, with the modulation time reference set by the second photon of the pair~\cite{Kolchin2008}.

Here, we instead vary the radiative relaxation time of a superconducting qubit throughout the relaxation process by controlling the coupling strength to a microwave coplanar waveguide cavity~\cite{Houck2007,Filipp2011}. The qubit-cavity coupling influencing the relaxation rate is known as the Purcell effect~\cite{Purcell1946}. Dynamically tunable coupling is possible with the tunable coupling qubit (TCQ)~\cite{Srinivasan2011}.

The TCQ is a single superconducting charge qubit internally composed of two directly coupled transmons. It maintains the transmon's favorable properties including charge noise insensitivity~\cite{Gambetta2011} and the ability to tune the individual transmons' energy levels in tens of nanoseconds using fast flux bias lines~\cite{Hoffman2011}. The latter feature leads to independent control over both the TCQ's frequency and coupling. This independence is essential to achieving high fidelity quantum state transfer. In order to produce a time-symmetric wave packet, not only must the envelope be symmetric, but the dressed qubit frequency during the emission process must be held constant. Changes in the emission frequency correspond to the Bloch vector's polar angle $\phi$ shifting during the emission process and produce phase errors which ruin the time-reversal symmetry.

The device used here consists of a TCQ coupled to a niobium superconducting coplanar waveguide cavity. The input port of the cavity is weakly coupled and is used to prepare the qubit state. The output port is coupled strongly to the amplifier chain and therefore radiative emission from the qubit exits the system primarily through the output port. This asymmetric cavity has a linewidth of $\kappa = 20~\unit{MHz}$ and a bare resonance at $\omega_\mathrm{c} = 7.596~\unit{GHz}$. The maximum TCQ frequency is $\omega_{\mathrm{TCQ}}^{\mathrm{max}} = 8.741~\unit{GHz}$. During the pulse shaping experiment, the TCQ is operated at a constant dressed frequency of $\omega_{\mathrm{TCQ}} = 7.445~\unit{GHz}$.  The two flux bias lines change the TCQ's coupling strength $g(t)$ while maintaining a constant dressed frequency~\cite{Hoffman2011}. The sample is cooled in a dilution refrigerator and care is taken to shield the sample from thermal radiation~\cite{Corcoles2011,Barends2011}.

In the Purcell dominated regime, the TCQ is an on-demand single photon source~\cite{Houck2007}. The operating point where the qubit state is initialized is chosen to have an intermediate value of $T_1=470~\unit{ns}$. If the qubit's state is prepared in a regime with low $T_1$, then the system decays before the coupling can be varied appropriately to generate the symmetric shape. On the other hand, initialization in regions of high $T_1$ and small coupling requires a strong drive. This can populate the cavity significantly since the qubit is close to the cavity resonance, leading to decoherence. Emitted photons are measured by mixing the cavity signal with a local oscillator at the emission frequency and digitizing both quadratures of the DC signal with a high speed acquisition card.

\begin{figure}
\begin{center}
	\includegraphics[width = .7\textwidth]{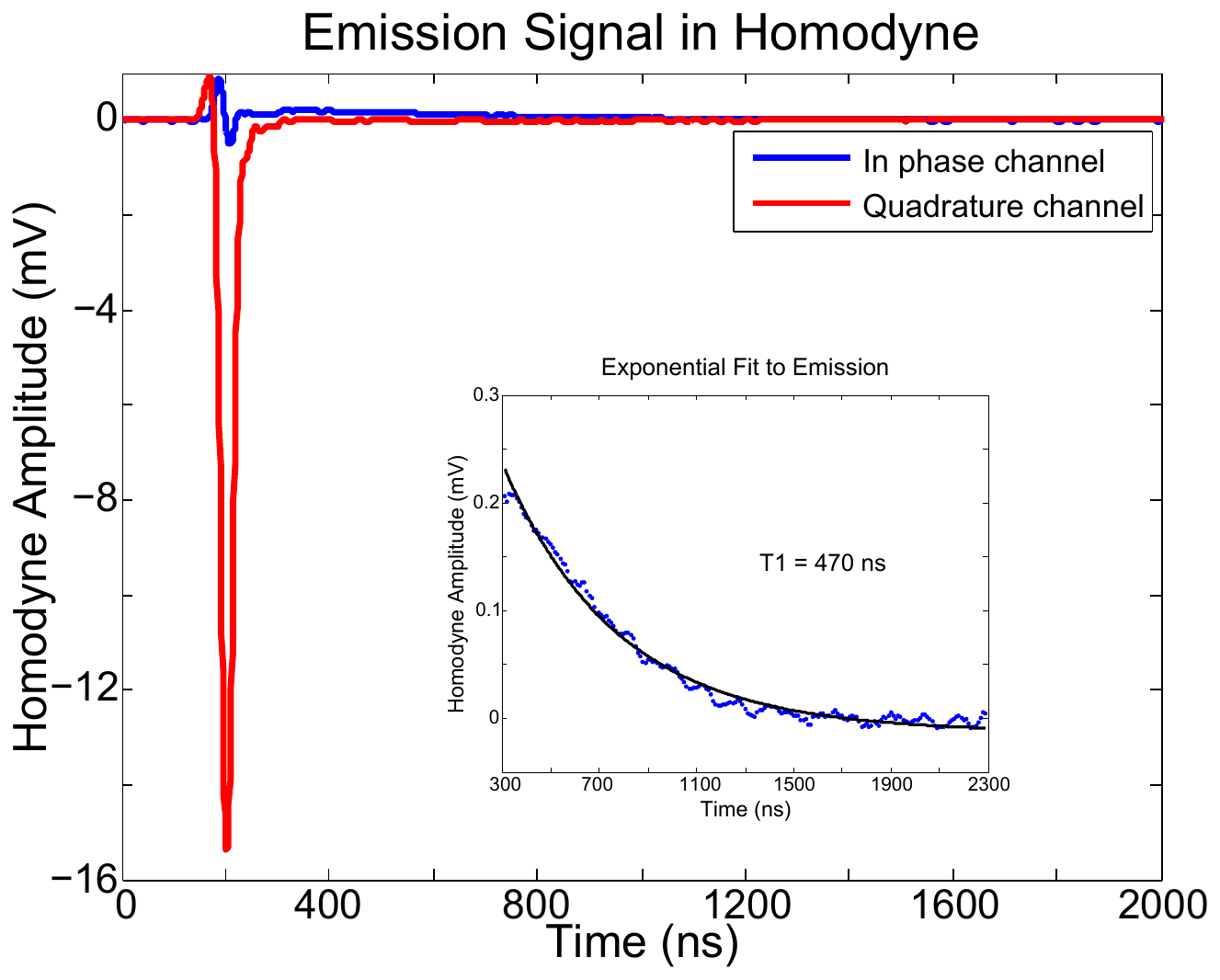}
\caption{\label{labStaticEmission}Radiative Relaxation at Initialization Point. With the bare cavity frequency at $\omega_\mathrm{c} = 7.596~\unit{GHz}$ and linewidth of $\kappa = 20~\unit{MHz}$, the TCQ frequency is fixed at $\omega_\mathrm{TCQ} = 7.445~\unit{GHz}$. A $\pi/2$ pulse is applied at $\omega_\mathrm{TCQ}$ to prepare the TCQ in the maximal superposition state, and homodyne voltages during relaxation are measured by amplifying and mixing the cavity output with a local oscillator also at $\omega_\mathrm{TCQ}$. The local oscillator phase is adjusted such that the excitation pulse, visible between 150 ns and 250 ns, is primarily in the quadrature channel of the measurement. This corresponds to performing a Y-rotation on the Bloch sphere, which puts the qubit Bloch vector on the equatorial plane parallel with the X-axis. The photon is emitted by the qubit along the quadrature orthogonal to the excitation direction, and is therefore measured in the in-phase channel. The amplitude of the drive pulse is considerably larger than the amplitude of photon emission which spreads over a large time window; therefore the emission is shown by zooming into the signal for time greater than 300 ns in the inset. The symmetric pulse in Fig.~\ref{labSymShape} is generated by preparing the superposition state similar to the above experiment, but tuning the coupling strength according to Eq.~\ref{eq_gt}.}
\end{center}
\end{figure}

Since homodyne detection measures the phase of the electromagnetic field, it cannot be used to detect a state of light that has maximal phase uncertainty, such as Fock states containing a well defined number of photons. Hence a $\pi/2$ pulse is used to prepare the qubit on the equatorial plane of the Bloch sphere, which upon relaxation emits the $(\ket{0} + \ket{1})/\sqrt{2}$ photon state. The phase of the excitation pulse is tuned such that it lies entirely in the quadrature channel as a large spike at $200~\unit{ns}$ in Fig.~\ref{labStaticEmission}. The in-phase and quadrature channels of the excitation pulse produce qubit rotations around the $X$ and $Y$ axes, respectively. Qubit excitations around the $Y$ axis (quadrature channel) produce single-photon homodyne signal along the $X$ axis (in-phase channel) and therefore a natural separation exists between the drive and emission signals~\cite{Houck2007}. The inset in Fig.~\ref{labStaticEmission} shows the radiative relaxation in the in-phase channel after the excitation pulse in the quadrature channel. The pulse shaping experiment described in the following paragraphs starts with an identical initialization, after which the qubit-cavity coupling strength $g(t)$ is appropriately tuned to generate the desired wave packet.

To determine the exact form for $g(t)$, consider first the photon flux $Q(t)$ exiting the cavity. Within a Markov approximation, $Q(t)$ can be related to the probability $P(t)$ of finding the TCQ in the excited state by
\begin{equation}
  -\frac{dP(t)}{dt} \:=\: Q(t) \:=\:\frac{4 g^2(t)}{\kappa} P(t) \: \: .
\end{equation}
Integrating the first pair of equations, using the initial condition that at $t=-\infty$ the qubit has unit probability of being in the first excited state, and then solving the second pair describing the Purcell decay~\cite{Purcell1946} for $g^2(t)$, we arrive at
\begin{equation}
\label{gsquared}
  g^2(t) \:=\: \frac{\kappa}{4}
  \frac{Q(t)}{1-\int_{-\infty}^t Q(t^{'}) \ dt^{'}} \: \: .
\end{equation}
Symmetrizing the emitted photon packet amplitude requires symmetrizing the photon flux. The simplest symmetric shape is the symmetric exponential, where $Q(t) = \frac{\beta}{2} e^{-\beta |t|}$, with $\beta$ some time constant. Using this symmetric pulse in \eqref{gsquared} gives a time-dependent coupling
\begin{eqnarray}\label{eq_gt}
  g^2(t) \:=\:
  \left \{
  \begin{array}{crl}     
    \frac{\kappa \beta}{4} \frac{\frac{1}{2} e^{-\beta |t|}}{1-\frac{1}{2} e^{-\beta |t|}}
      & \ \ \ \ \ t \le 0 & \\
    \frac{\kappa \beta}{4}
      & \ \ \ \ \ t > 0 & .
  \end{array}
  \right .
\end{eqnarray}
A symmetric photon flux requires tuning the coupling in an \emph{asymmetric} way. Rather than directly varying the coupling, we reformulate the problem into one of tuning the radiative relaxation time, which can be measured more easily. In terms of the coupling strength, it is $T_1(t)=\frac{\kappa}{4 g^2(t)}$. One advantage of using this wave packet shape is that the falling half does not require tuning of $T_1$, since holding $T_1$ constant produces an exponential decay, as depicted in Fig.~\ref{labStaticEmission}. Another advantage is that this technique ensures full decay of the excited state population, which otherwise could lead to a reduction in the state transfer fidelity. A disadvantage of this pulse shape is that it requires truncating the rising side of the wave packet due to the finite maximum $T_1$ the TCQ can achieve. However, the maximum $T_1$ can be increased by improved fabrication techniques~\cite{Sandberg,Chang}.

\begin{figure}
\begin{center}
	\includegraphics[width=.57\textwidth]{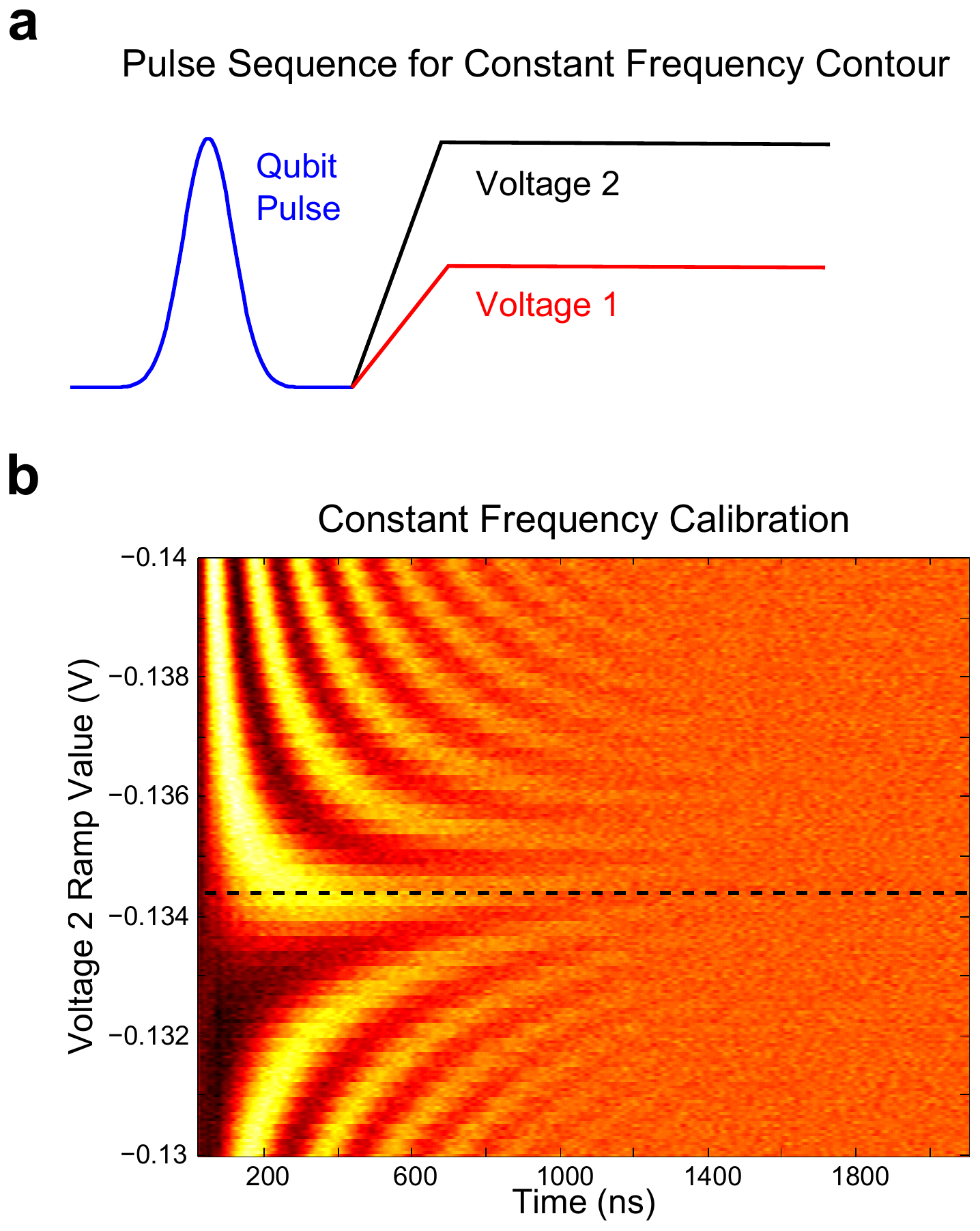}
\caption{\label{labPulse_Sweep}Using Single Photon Emission to Calibrate Constant Frequency Contour. a) First, initialization is performed by a $\pi/2$ qubit pulse (blue) resonant at the qubit transition frequency. This prepares the $\Big(\ket{0} + \ket{1}\Big)/\sqrt{2}$ qubit state. Before the relaxation happens, the qubit-cavity coupling strength and transition energy are varied by applying two voltage pulses of varying magnitude (red and black) to the two SQUID loops of the TCQ. b) The emission signal is measured by homodyne and plotted as a function of time and Voltage 2 amplitude, as Voltage 1 is kept constant. The fringes in the time axis are due to the emission occurring at a frequency different from the initialized qubit frequency. There is a Voltage 2 for every Voltage 1 where the fringes in the emission are minimized, which in this plot occurs at a Voltage 2 value of $-0.1345 \unit{V}$, indicated by the dotted line. The decay of the emission signal is a direct measure of the radiative relaxation rate, $T_1$ (which is a function of the qubit-cavity coupling strength). The line of constant phase tends upwards for times greater than $1000~\unit{ns}$; this is due to imperfections in the pulse programmer used for the fast voltage pulses. Fortunately, the symmetric pulse shaping experiment is immune to this problem since the emission pulse is engineered to decay well before such imperfections occur.}
\end{center}
\end{figure}

While the radiative $T_1(t)$ is varied, the dressed emission frequency must remain constant. Varying the emission frequency causes precession of the qubit state vector around the z-axis of the Bloch sphere, with respect to the rotating frame defined by the local oscillator of the measurement. This leads to oscillations in both quadratures of the homodyne voltage. These oscillations are used to calibrate a constant frequency contour in the space spanned by the two flux-line voltages that control the TCQ's energy levels. The pulse sequence used for this calibration is shown in Fig.~\ref{labPulse_Sweep}a, with the $\pi/2$ pulse used to excite the qubit shown in blue, and red and black representing the two voltage pulses used to change the TCQ's energy levels on short time scales. Despite both the TCQ's coupling and frequency changing as a result of these voltage pulses, the excitation power required to initialize the qubit to the full superposition state is the same since the excitation always occurs when the voltage pulses are zero. One can find combinations of the two voltages that produce constant emission frequency by holding one voltage constant and sweeping the other as in Fig.~\ref{labPulse_Sweep}b. The emission signal takes a chevron shape with the peak corresponding to constant frequency emission. The fringes away from the central peak oscillate at a frequency equal to the detuning, similar to a Ramsey interference measurement.

\begin{figure}
\begin{center}
	\includegraphics[width=.7\textwidth]{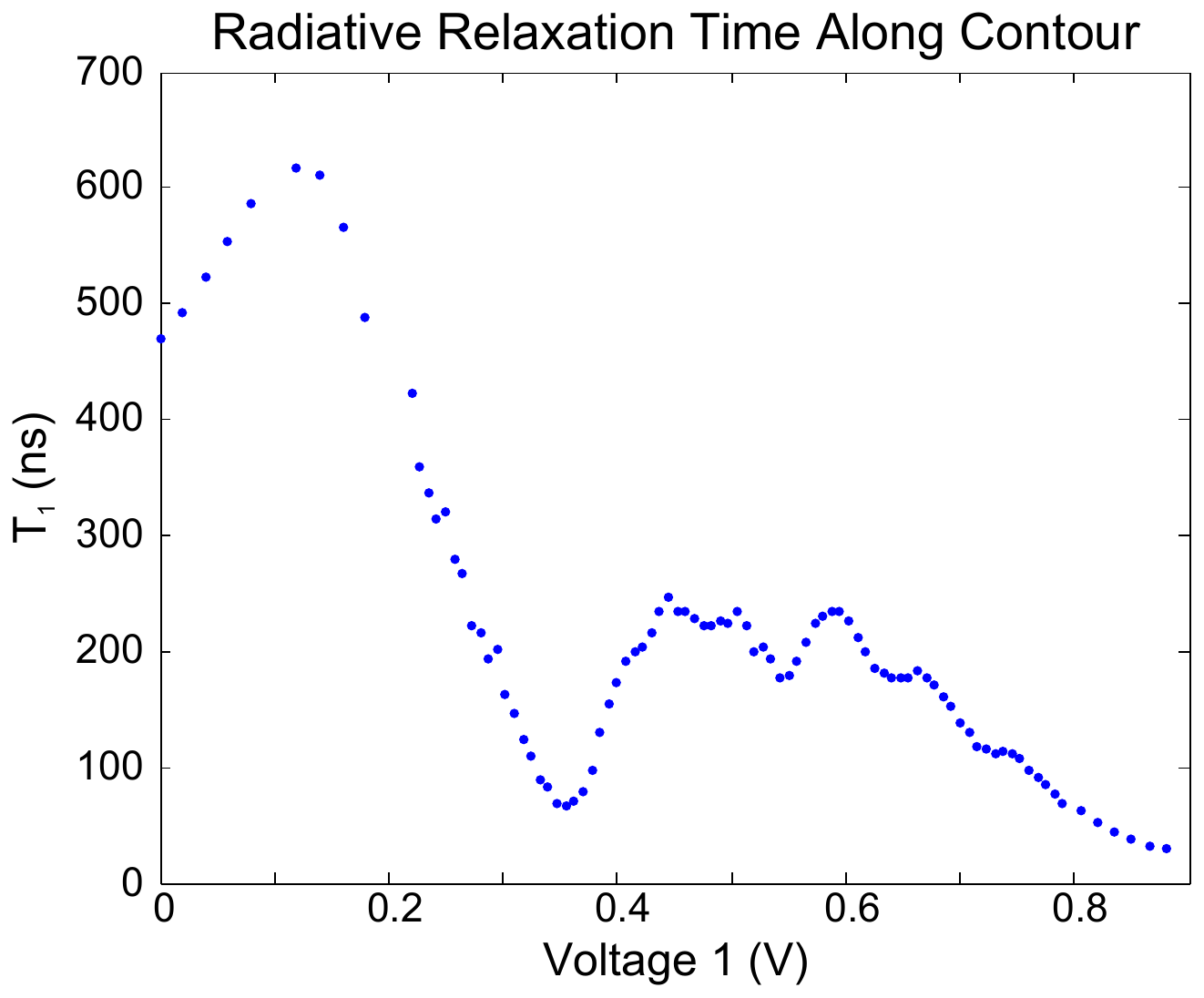}
\caption{\label{labT1}Radiative Relaxation Time. Once the contour of constant frequency has been mapped using the pulsing and emission measurements described in Fig.~\ref{labPulse_Sweep}, the resonant emission signal is extracted and fitted with an exponential decay, with the decay constant equal to the radiative emission time, $T_1$. The relaxation time is not monotonic because of fabrication mismatches in the geometry of the left and right halves of the TCQ. The maximum radiative relaxation rate measured is just above $600~\unit{ns}$.}
\end{center}
\end{figure}

\begin{figure}
\begin{center}
	\includegraphics[width=.8\textwidth]{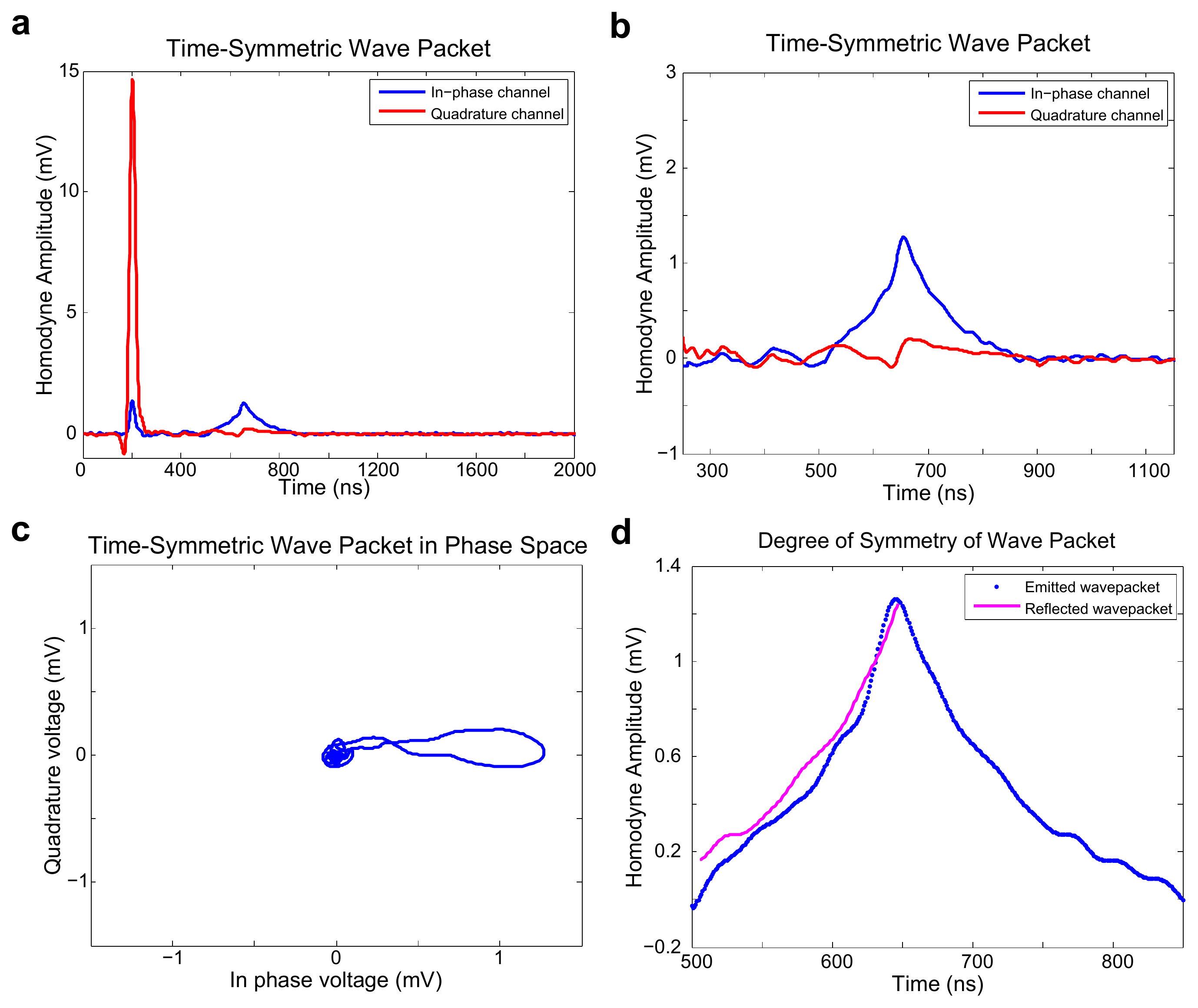}
\caption{\label{labSymShape}Time-symmetric wave packet for Quantum State Transfer. a-b) The qubit is initialized with a $\pi/2$ pulse at an operating point where $T_1 = 470~\unit{ns}$, similar to the static emission measurement in Fig.~\ref{labStaticEmission}. The qubit excitation pulse is again contained between 150 ns and 250 ns. But now as the qubit relaxes, $T_1$ is tuned to generate the symmetric exponential wave packet. Since the photon flux (and hence the wave packet shape) is directly proportional to the excited state population, the pulse rise is generated by reducing $T_1$ asymmetrically to compensate for the ever decreasing excited state population. The falling half of the pulse is generated by holding $T_1$ constant, which produces an exponentially decaying wave packet. The symmetric emission lies almost entirely in the in-phase channel. This is a signature of minimal frequency drift during the emission process, which is essential for efficient state transfer to a second cavity. In b), the plot focuses on the time during emission pulse. c) To quantify the frequency drift, the emission is plotted in phase space. A truly constant frequency signal would have minimal width along the quadrature axis. d) To view the degree of symmetry of the wave packet, the falling half (magenta) is superimposed onto the pulse rise (blue) for comparison. Due to finite relaxation times, the pulse rise needs to be truncated and hence the emission starts at $500~\unit{ns}$.}
\end{center}
\end{figure}

The resonant emission decays exponentially with a decay constant equal to the radiative relaxation time. This $T_1$ is plotted in Fig.~\ref{labT1} along the contour of constant frequency. $T_1$ can be tuned from $600~\unit{ns}$, where the coupling strength is weakest, to less than $50~\unit{ns}$, where $T_1$ is dominated by the cavity linewidth $\kappa$. To generate a symmetric pulse shape, $T_1$ is decreased from its maximum value along this contour during the rising edge of the pulse, and then held constant at $50~\unit{ns}$ to generate the exponential for the falling half. The photon pulse shape following Eq.~\ref{eq_gt} is shown in Fig.~\ref{labSymShape}a. In the time interval spanning $150$ to $250~\unit{ns}$, the $\pi/2$ excitation pulse is applied, and the measurement axis is aligned such that the pulse is in the quadrature channel. Once the excitation is applied, the qubit is moved to a region of high $T_1$ to add some separation in time between the excitation and the emission pulses. Following this waiting period, the $T_1$ is again varied to produce the desired pulse shape. The emission lies primarily in the in-phase channel, but some residual emission is seen in the quadrature channel. This corresponds to small drifts in the frequency during the emission process. 


Alternatively, one can view the emission in the I/Q phase space as shown in Fig.~\ref{labSymShape}c. Constant frequency emission corresponds to a straight line, wherein the phase remains constant throughout the emission process. In Fig.~\ref{labSymShape}c, the emission is seen to be mostly parallel to the in-phase axis, with a drift of about 5 degrees. The degree of amplitude symmetry is measured by fitting an exponential to each half of the pulse: the rising half is slightly faster than the falling half, with time constants of $45~\unit{ns}$ and $49~\unit{ns}$, respectively, and both containing the $50~\unit{ns}$ target within the $95\%$ confidence interval of the fit. The mismatch between the time constants is due in large part to the truncation required on the rising half. To visually compare the rising half and falling half, the latter is reflected and superimposed onto the former in Fig.~\ref{labSymShape}d.

\begin{figure}
\begin{center}
	\includegraphics[width=.95\textwidth]{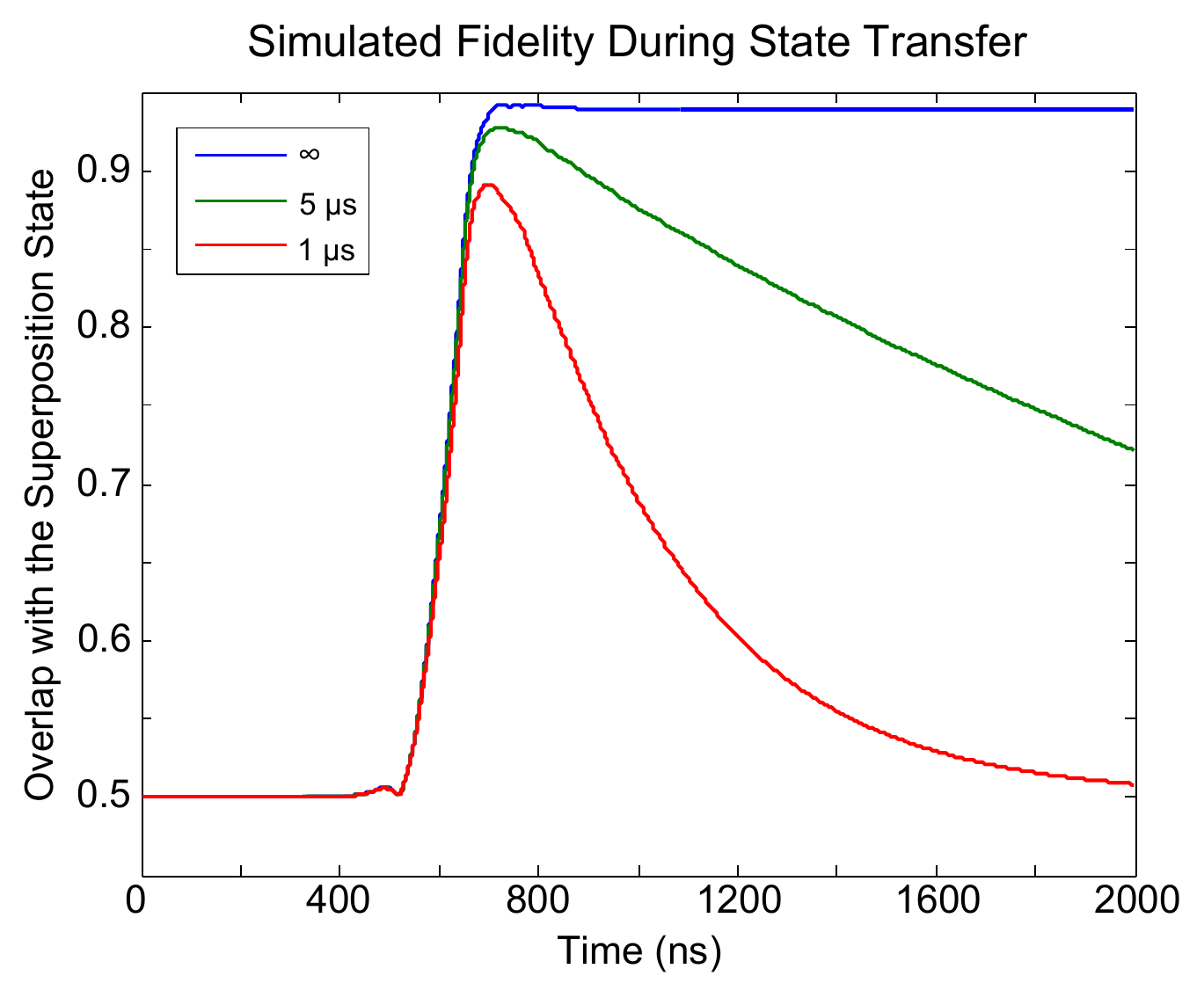}
	\caption{\label{labCatch}Simulating State Transfer. The state transfer fidelity, $F(t) = \mathrm{Tr}\big(\rho(t) \hat{F}\big)$, is plotted for the case where the experimentally produced wave packet in Fig.~\ref{labSymShape} is sent towards a TCQ-cavity system. Here, $\hat{F}$ is the fidelity operator which is the tensor product of an identity on the photons and the projector onto the target superposition state. The TCQ is initially in the ground state and its coupling strength to the cavity is tuned in a manner time-reversed to the form described in Eq.~\ref{eq_gt}, which was used to generate the symmetric shape. Since the TCQ in the destination cavity is initially in the ground state, the overlap fidelity starts at $0.5$ before the wave packet drives it into the symmetric superposition. Effects of non-radiative relaxation are considered for values $T_\perp = 1.0~\unit{\mu s}$ (red), $5~\unit{\mu s}$ (green), and $\infty$ (blue). With $T_\perp = \infty$, the maximum fidelity is $94\%$.}
\end{center}
\end{figure}

To understand how well this experimentally generated wave packet would perform in a state transfer experiment, quantum Monte Carlo simulations are used to model the absorption of this wave packet by another TCQ in a separate cavity. Note that neither the second TCQ nor the second cavity have to be strictly identical to the first.  One only needs enough control freedom in the dressed frequency and dressed decay rate of the TCQ-cavity combination to match the two systems. In the simulation, both quadratures of the wave packet shown in Fig.~\ref{labSymShape}, including the asymmetry and truncation errors, are used to drive a resonant TCQ which is initially in the ground state. The coupling strength used, ${\bar g}(t)=g(T_0 - t)$, is the time-reverse of the form in Eq.~\ref{eq_gt} and $T_0$ accounts for the light travel time between the cavities. The experimentally generated wave packet encodes the symmetric superposition state from the source TCQ-cavity system. Fig.~\ref{labCatch} presents the time evolution of the fidelity $F(t)=\mathrm{Tr}\big(\rho(t) \hat{F}\big)$~\cite{Nielsen2000} during the wave packet absorption, with $\rho(t)$ the density matrix of the destination TCQ-cavity system and the fidelity operator defined as the tensor product of the identity operator on the cavity photons ($\hat{I}$) and the projector onto the target symmetric qubit state $\ket{+}=\frac{1}{\sqrt{2}}(\ket{g}+\ket{e})$. Therefore, the fidelity operator is $\hat{F}=\hat{I} \otimes \ket{+}\bra{+}$. At the destination TCQ, non-radiative relaxation times of $T_\perp = \infty$, $5~\unit{\mu s}$, and $1~\unit{\mu s}$ are considered. With $T_\perp = \infty$ the fidelity reaches a maximum of $94\%$, with the remaining losses coming from the imperfections in the experimentally generated waveform. The sample used for generating the symmetric wave packet in this work had $T_\perp = 1.6~\unit{\mu s}$. Adopting improved fabrication techniques will help prolong $T_\perp$ and enable quantum state transfer~\cite{Chang}.

In this work, the qubit-cavity coupling strength in a circuit QED system is tuned on time scales much shorter than the coherence time. Operating the TCQ in the Purcell dominated regime enables precise control over the radiative relaxation rate. This feature is used to spontaneously emit a photon with a symmetric exponential wave packet. Such a time-symmetric wave packet can be fed into another TCQ-cavity system to realize quantum state transfer between qubits in different cavities. Finally, simulations are used to model the fidelity for the case where the actual experimentally generated wave packet is used in a state transfer experiment.


\begin{acknowledgments}
This work was supported by IARPA under Contract W911NF-10-1-0324 and ARO under Contract W911NF-11-1-0086. SMG would like to acknowledge NSF DMR-1004406 and ARO W911NF-05-1-0365.
\end{acknowledgments}

\clearpage

\clearpage

\clearpage

\clearpage

\clearpage

\clearpage

\clearpage
\bibliographystyle{apsrev4-1}
\bibliography{tunablecoupling}
\end{document}